\documentstyle[12pt]{article}
\begin{document}
\title{ON THE GRAVITOMAGNETIC EFFECTS IN CYLINDRICALLY SYMMETRIC
SPACETIMES}
\author{L. Herrera$^1$\thanks{Postal
address: Apartado 80793, Caracas 1080A, Venezuela.} \thanks{e-mail:
laherrera@telcel.net.ve}
 and N. O.
Santos$^2$\thanks{e-mail: nos@cbpf.br}
\\ \\
{\small $^1$Area de F\'{\i}sica  Te\'orica, Facultad de Ciencias}\\
{\small Universidad de Salamanca, 37008 Salamanca, Spain}\\
{\small $^2$Laborat\'orio Nacional de Computa\c{c}\~ao Cient\'{\i}fica,}\\
{\small 25651-070 Petr\'opolis RJ, Brazil, and}\\
{\small LAFEX, Centro Brasileiro de Pesquisas F\'{\i}sicas,}\\
{\small rua Dr. Xavier Sigaud 150, Urca,}\\
{\small 22290-180 Rio de Janeiro RJ, Brazil.}}
\maketitle
\begin{abstract}
Using gyroscopes we generalize results, obtained for the gravitomagnetic
clock effect in the particular case when the exterior spacetime is produced
by a rotating dust cylinder, to the case when the vacuum spacetime is
described by the general cylindrically symmetric Lewis spacetime. Results
are contrasted
with those obtained for the Kerr spacetime.
\end{abstract}
\section{Introduction}
Recently Bonnor and Steadman \cite{Bonnor} calculated and analysed the
gravitomagnetic clock effect, which is the difference in periods of a test
particle moving in prograde and retrogade circular geodesic orbits around
the axis of a rotating body. They applied their results to a cylindrically
symmetric system produced by van Stockum metric \cite{Stockum} describing a
rotating dust cylinder. The exterior spacetime, containing two parameters,
is a particular case of the general vacuum stationary cylindrically
symmetric Lewis metric \cite{Stockum,Lewis,Kramer} containing four
parameters. We extend some of their results to the general Lewis spacetime
by using the results obtained by us \cite{Herrera} for the gyroscope
precession in cylindrically symmetric spacetimes. The clock effect and the
gyroscope precession amount to similar physical processes. However, as it
will be seen below, using gyroscopes allows for a wider class of possible
"gedanken" experiments. Indeed, we have to face with two different effects:
one is the influence of the rotation of the source on the gravitational
field where the gyroscope is placed (the gravitomagnetic effect), which of
course is absent in Newtonian theory; the other is related with the fact
that the frame of the gyroscope may be rotating, producing a precession in
the gyroscope (Thomas-like precession).
\section{Precession of a gyroscope moving in a circle around the axis of
symmetry}
The Lewis metric can be written as
\begin{equation}
ds^2=-fdt^2+2kdtd\phi+e^{\mu}(dr^2+dz^2)+ld\phi^2,
\end{equation}
where
\begin{eqnarray}
f=ar^{1-n}-\frac{c^2r^{1+n}}{n^2a},\\
k=-Af,\\
l=\frac{r^2}{f}-A^2f,\\
e^{\mu}=r^{(n^2-1)/2},
\end{eqnarray}
with
\begin{equation}
A=\frac{cr^{1+n}}{naf}+b.
\end{equation}
The parameters $n, a, b$ and $c$ can be either real or complex, and
the corresponding solutions belong to the Weyl or Lewis classes
respectively. Here we restrict our study to the Weyl class (not to confound
with Weyl metrics representing static and axially symmetric spacetimes).

The parameters $n$ and $a$ are proportional to the Newtonian energy per
unit length and
the topological defect, respectively; while $b$ and $c$ describe the
stationarity of the source and are proportional to the angular momentum of
the
source producing a topological defect and the vorticity of the source,
respectively.

Now it is important to stress that the transformations \cite{Stachel}
\begin{eqnarray}
d\tau=\sqrt{a}(dt+bd\phi),\\
d\bar{\phi}=\frac{1}{n}[-cdt+(n-bc)d\phi],
\end{eqnarray}
cast the Weyl class of the Lewis metric into the Levi-Civita cylindrical
metric
(static). However the transformations above are not valid globally, and
therefore both metrics are equivalent only locally, a fact that can be
verified by calculating the corresponding Cartan scalars \cite{Silva}. In
order to globally transform the Weyl class of the Lewis metric into the
static Levi-Civita metric, we have to make $b=0$. Indeed, if $b=0$ and $c$
is different from zero, (7) gives an admissible transformation for the time
coordinate and (8) represents the transformation to a rotating frame
(implying thereby that the frame of (1) is itself rotating). In other
words, if $b=0$, (1) is just the exterior line element of a static cylinder,
as seen by a rotating observer. However, since rotating frames (as in
special relativity) are not expected to cover the whole spacetime, and
furthermore since the new angle coordinate ranges from $-\infty$ to
$\infty$, it has been argued in the past \cite{Silva} that both $b$ and $c$
have to vanish for (7) and (8) to be globally valid. This point of view is
also reinforced by the fact that, assuming that only $b$ has to vanish in
order to globally cast (1) into Levi-Civita, we are lead to the intriguing
result that there is no dragging outside rotating cylinders. We shall recall
this question later.

The rotation $\Omega$ of the compass of inertia, or the gyroscope, with
respect to a rotating frame with angular velocity $\omega$ moving around the
axis of symmetry given by metric (1) can be easily calculated by using the
Rindler-Perlick method \cite{Rindler}.

This consists in transforming the angular coordinate $\phi$ by
\begin{equation}
\phi=\phi^{\prime}+\omega t,
\end{equation}
where $\omega$ is a constant (observe that (8), with $b=0$, defines a
rotation in the sense opposite to that in (9)). Then the transformed metric
is written in a
canonical form
\begin{equation}
ds^2=-e^{2\Psi}(dt-\omega_idx^i)^2+h_{ij}dx^idx^j,
\end{equation}
with latin indexes running from 1 to 3 and $\Psi, \omega_i$ and $h_{ij}$
depend on the spatial coordinate $x^i$ only (we are omitting primes). Then,
it may be shown that the four-acceleration $A_{\mu}$ and the rotation
three-vector
$\Omega^i$ of the congruence of world lines $x^i=$ constant are given
by
\begin{eqnarray}
A_{\mu}=(0,\Psi_{,i}),\\
\Omega^i=\frac{1}{2}e^{\Psi}(\det h_{mn})^{-1/2}\epsilon^{ijk}\omega_{k,j},
\end{eqnarray}
where the comma denotes partial derivative. It is clear from the above that
if $\Psi_{,i}=0$, then particles at rest in the rotating frame follow
circular geodesics. On the other hand, since $\Omega^i$ describes the rate
of rotation with respect to the proper time at any point at rest in the
rotating frame, relative to the local compass of inertia, then $-\Omega^i$
describes the rotation of the compass of inertia (the {\it gyroscope}) with
respect to the rotating frame. Applying (9) to the original frame of (1),
with $t=t^{\prime}, r=r^{\prime}$ and $z=z^{\prime}$, we cast (1) into the
canonical form (10), and obtain (see (43) in \cite{Herrera})
\begin{equation}
\Omega=MNr^{(1-n^2)/4}\left(M^2ar^{1-n}-\frac{N^2r^{1+n}}{n^2a}\right)^{-1},
\end{equation}
where
\begin{equation}
M=1+b\omega, \;\; N=n\omega-c(1+b\omega).
\end{equation}
{}From (13) we can ask if there are $\omega$'s for which the gyroscope
precession is null.

We see from (13) that the gyroscope does not precess if $M=0$ or $N=0$, thus
producing $\Omega=0$ and implying respectively for the angular velocity of
the frame
\begin{equation}
\omega_M=-\frac{1}{b}, \;\;\omega_N=\frac{c}{n-bc}.
\end{equation}
The physical meaning of this result will be discussed below. A similar
result has been obtained in \cite{Bonnor} but in the particular context of
van Stockum solution, while our result is general and independent of the
source.

The tangential velocity $W$ of the gyroscope moving around the axis of
symmetry for metric (1) is given by (see (53) in \cite{Herrera1})
\begin{equation}
W=\frac{\omega(fl+k^2)^{1/2}}{f-\omega k}.
\end{equation}
Substituting (2-5) into (16), we obtain
\begin{equation}
W=\frac{n\omega\chi}{(1+b\omega)(1-c^2\chi^2)+nc\omega\chi^2},
\end{equation}
where
\begin{equation}
\chi=\frac{r^n}{na}.
\end{equation}
The angular velocities (15) give, respectively, from (17), the tangential
velocities
\begin{equation}
W_M=\frac{1}{c\chi}, \;\; W_N=c\chi,
\end{equation}
and we observe that these velocities do not depend upon $b$ in spite of the
corresponding angular velocities depend upon $b$.

The Newtonian energy per unit length $\sigma$ is given, in terms of $n$, by
\begin{equation}
\sigma=\frac{1}{4}(1-n),
\end{equation}
and we consider the range $1>n>-1$ or $0<\sigma<1/2$. This range produces
physically reasonable cylindrically symmetric sources \cite{Silva}. However
there exist no circular timelike geodesics for $n<0$, and furthermore it is
not
clear that $n<0$ represent cylinders \cite{jorge}.

{}From (19) we see that as $r\rightarrow 0$, for $1>n>0$,
$W_{M+}\rightarrow\infty$ and $W_{N+}\rightarrow 0$; while for $0>n>-1$,
$W_{M-}\rightarrow 0$ and $W_{N-}\rightarrow -\infty$. We discard $W_{M+}$
and $W_{N-}$ as being unphysical.

Now, let us suppose that $1>n>0$, then $\Omega$ vanishes for
$\omega=\omega_N$. If furthermore $b=0$, then it follows at once from (8),
that transformation (9) brings the system back to the non-rotating frame
(the frame in which the line element is static), thereby explaining the
vanishing of the precession. The remarkable fact, however, is that $\Omega$
vanishes for $\omega_N$, even if $b$ is different from zero. As for
$\omega_M$, we have not a reasonable interpretation, unless we accept that
(1) describes a cylinder only if $1>n>0$.

Now we study the case of infinite precession, $\Omega\rightarrow\infty$, for
the gyroscope moving around the axis of symmetry. From (13) we have then
\begin{equation}
r^n=\frac{Mna}{N},
\end{equation}
and considering (14), we can rewrite (21) for the angular velocity of the
rotating frame,
\begin{equation}
\omega=\frac{1+c\chi}{n\chi-b(1+c\chi)}.
\end{equation}
The corresponding tangential speed of the gyroscope becomes, using (16),
(17) and (22),
\begin{equation}
W=1,
\end{equation}
which means that the gyroscope attains infinite precession when its
tangential velocity around the axis becomes the light velocity.

\section{Precession of a gyroscope at rest}
If the gyroscope is at rest in the original lattice, then we have (see (32)
in \cite{Herrera})
\begin{equation}
\Omega=\frac{cr^{(1-n)(n-3)/4}}{a(1-c^2\chi^2)}.
\end{equation}
Observe that it is the absolute value of $\Omega$ what appears in
(31),(32),(33) and (34) in \cite{Herrera}.
We see that the precession is infinite if $c\chi=1$. It is remarkable that
for $c\chi=1$, if the gyroscope is moving around the axis of symmetry,
produces a tangential speed of light (19), $W_{N+}=1$, with null precession;
on the other hand, in this same case $c\chi=1$, while at rest its precession
becomes infinite.

On the other hand, when $b=0$
and $c=0$, i.e., when the Weyl class of Lewis metric becomes the static
Levi-Civita cylindrical
spacetime, the precession of a gyroscope moving around the axis of
symmetry results in
\begin{equation}
\Omega=\frac{n\omega r^{(1-n)(n-3)/4}}{a(1-n^2\omega^2\chi^2)},
\end{equation}
with a tangential velocity obtained from (17)
\begin{equation}
W=n\omega\chi.
\end{equation}
We observe that the gyroscope precession is the same in both cases, (24) and
(25), if the angular velocity of the gyroscope, in the Levi-Civita
spacetime, is related to the vorticity of Lewis spacetime by
\begin{equation}
\omega=\frac{c}{n}.
\end{equation}
These two equal precessions, (24) and (25), suggest that (if $b=0$) it is
equivalent to
measure the precession of a gyroscope at rest with respect to the rotating
Lewis source or moving around the corresponding static source. This
situation, in
turn, is a reminiscense of the non-Machian behaviour of Newtonian gravity,
where gravitomagnetic effects are absent.

\section{Precession of a gyroscope in a locally non rotating frame}
Using the transformation
\begin{equation}
d\phi=d\bar{\phi}+\omega dt,
\end{equation}
where $\omega$ is
\begin{equation}
\omega=-\frac{k}{l},
\end{equation}
the Lewis metric (1) transforms into a diagonal form near $r=r_0$. This
frame is called locally non-rotating \cite{Bonnor1,Silva1}. From (28) for
the Lewis metric (1) we have
\begin{equation}
\omega=\frac{n^3a^2c-2n^2a^2bc^2+(bc-n)c^3r_0^{2n}+n^4a^4br_0^{-2n}}{ n^4a^2
-2n^3a^2bc+2n^2a^2b^2c^2-(n-bc)c^2r_0^{2n}-n^4a^4b^2r_0^{-2n}},
\end{equation}
which can be rewritten with (18),
\begin{equation}
\omega=\frac{(n-bc)c\chi^2_0+b}{(n-bc)^2\chi^2_0-b^2},
\end{equation}
where $\chi_0=\chi(r_0)$. The tangential velocity (17) with (31) becomes
\begin{equation}
W=\frac{(n-bc)c\chi^2_0+b}{n\chi_0},
\end{equation}
and the precession (13) with (31) becomes
\begin{equation}
\Omega=\frac{b(n-bc)r_0^{(1-n)(n-3)/4}}{a[(n-bc)^2\chi^2_0-b^2]}.
\end{equation}
{}From (31) we see that there are two cases where $\omega$ does not depend upon a
particular radius $r_0$ and produces no precession according to (33). These
cases are, for $b=0$,
\begin{equation}
\omega=\frac{c}{n}, \;\; W=c\chi_0;
\end{equation}
and, for $bc=n$,
\begin{equation}
\omega=-\frac{1}{b}, \;\; W=\frac{1}{c\chi_0},
\end{equation}
where we have included, from (32), the corresponding tangential velocities.
We see from (34) that the result corresponds to what we obtained for
$\omega_N$ in (15) and agrees with the analysis
of the gyroscope at rest (24) compared to the precession in Levi-Civita's
spacetime (25). However the case (35), while producing a similar result
compared to $\omega_M$ in (15), imposes the relation $b=n/c$. When $b\neq 0$
and $b\neq n/c$ the locally non-rotating frame produces non-null precession.

\section{The Kerr spacetime}
It is instructive to  compare the situation described above with that in
the Kerr space.

In Boyer-Lindquist coordinates with $\theta=\pi/2$ the
Kerr metric has the form (the Kerr parameter $a$, describing angular
momentum per unit mass, not to be confounded with
the parameter $a$ of the Lewis metric)
\begin{eqnarray}
ds^2=-\left(1-\frac{2m}{r}\right)dt^2-\frac{4am}{r}dtd\phi+
\frac{1}{\Pi}dr^2 \nonumber \\
+\left(r^2+a^2+\frac{2a^2m}{r}\right)d\phi^2,
\end{eqnarray}
where
\begin{equation}
\Pi=1-\frac{2m}{r}+\frac{a^2}{r^2}.
\end{equation}
Then, applying the Rindler-Perlick method, one obtains after some lengthy
calculations
\begin{eqnarray}
e^{2\Psi}=\Lambda,\\
\omega_i=(0,0,\omega_{\phi}),\\
\omega_{\phi}=\frac{1}{\Lambda}\left[\omega(r^2+a^2)-\frac{2am}{r}(1-a\omega
) \right],\\
h_{rr}=\frac{1}{\Pi},\\
h_{\phi\phi}=\frac{\Pi}{\Lambda}r^2,
\end{eqnarray}
with
\begin{equation}
\Lambda=1-\omega^2(r^2+a^2)-\frac{2m}{r}(1-a\omega)^2.
\end{equation}

Substituting (38-42) into (12) we obtain
\begin{equation}
\Omega=\frac{2}{\Lambda}\left[\omega-\frac{3m}{r}\omega(1-a\omega)+\frac{am}
{r^3} (1-a\omega)^2\right].
\end{equation}
The value of the angular velocity $\omega$ for which there is no precession
($\Omega=0$) is easily obtained from
(44) to be
\begin{equation}
\omega= - \frac{r^2
(r-3m)-2ma^2-\sqrt{r^4(r-3m)^2-4ma^2r^3}}{2ma(3r^2+a^2)};
\end{equation}
this is the same value for which prograde and retrograde circular
geodesics have the same period \cite{semerak}, and which leads to
the condition of no clock effect in \cite{Bonnor}, after replacing
$\omega$ by its expression for a circular geodesic. This
result was obtained before \cite{felice} and (together with other
properties) led some authors to suggest that natural
non-rotating observers are those moving with angular velocity (45) (see
\cite{semerak} and references therein).
This identification, however, is not necessarily correct. In fact, observe
that a gyroscope at rest in the frame of (36)
($\omega=0$) will precess unless $a=0$, reflecting the well known fact that
the original frame of (36) is itself rotating with
 respect to a compass of inertia \cite{Rindler}. Therefore the vanishing of
$\Omega$ for observers rotating with angular velocity
 (45) only shows that the gravitational dragging effect of the source
exactly cancels the Thomas-like precession due to the rotation
 of the frame where the gyroscope is placed: a frame which, as shown in
\cite{Bonnor}, rotates relative to distant stars. Under these
circumstances it becomes difficult to accept that those observers represent
``the most natural standard of non-rotation"

\section{Conclusion}
We have seen that a gyroscope at rest in the frame of (1) will precess
independently of $b$, and in a similar way as a gyroscope moving around a
static source with angular velocity given by (27). This result, together
with
the fact that transformations (7) and (8) cast (1) into a static cylindrical
Levi-Civita's line element if $b=0$, would indicate that the rotation of the
source does not affect the gyroscope. However, for the gyroscope moving
around the source, there exist two possible angular velocities for which
there is no precession. The physical meaning of one of them ($\omega_M$) is
not
understood by the authors, unless the range of $n$ is restricted to $1>n>0$,
in which case it is discarded. The situation with $\omega_N$ is clear if
$b=0$, in which case (9) is just a transformation to the non rotating frame
if $\omega=\omega_N$. However, if $b$ is not vanishing then the reasons for
the vanishing of $\Omega$ are obscure. Finally if we define a locally non
rotating frame acording to (28) and (29) then we see that a gyroscope at
rest in such a frame will precess according to (33). The origin of this
precession is rather surprising if we note that it appears even if $n=a=1$
(Minkowski) and $c=0$. But under these conditions (1) is not the
Minkowskian line element corresponding to a rotating frame. So the question
here is, what is the nature of $b$, that makes the gyroscope to precess?

In the Kerr case we have seen that the frame in which $\Omega=0$ can hardly
be called non-rotating. The difference with Lewis case
(with $b=0$) becomes intelligible, if we note that the frame of (36) with
$m=0$ does not represent a rotating Minkowskian observer,
a conclusion confirmed by the fact that (44) with $m=\omega=0$ yields
$\Omega=0$. However, as mentioned before, the frame
of (36) is rotating with respect to a compass of inertia if $m\neq 0$
(yielding $\Omega\neq 0$). This is in contrast with the Lewis
case, where $\Omega$ is not vanishing for the gyroscope at rest in (1) in
the case $n=1$ (Minkowski). This conspicuous difference
in the relation between the source of the field and the rotation, in both
cases, seems to suggest, loosely speaking, that the
behaviour of the Kerr metric is more ``Machian'' than that of Lewis.

\end{document}